\documentclass[12pt]{article}
\usepackage{epsfig}
\usepackage{cite}
\begin{document}

\begin{flushright}
MAN/HEP/2003/01 \\
MC-TH-2003-01 \\
hep-ph/0303206\\
\end{flushright}

\bigskip

\begin{center}
{\bf {\LARGE Observing a Light CP-Violating Higgs Boson}}\\[3.5mm]
{\bf {\LARGE in Diffraction } }
\end{center}

\bigskip\bigskip

\begin{center}{\large 
B.E. Cox, J.R. Forshaw, J.S. Lee, J.W. Monk and A.~Pilaftsis }
\end{center}

\begin{center}{\em Department of Physics and Astronomy, University
of Manchester}\\
{\em Manchester M13 9PL, United Kingdom}\\
\end{center}

\bigskip\bigskip\bigskip

\centerline{\bf  Abstract} \noindent 
Light CP-violating Higgs bosons with mass lower than 70~GeV might have
escaped detection in  direct searches at  the LEP collider.  They  may
remain undetected in conventional  search channels at the Tevatron and
LHC.  In this Letter we  show that exclusive diffractive reactions
may be able to
probe for the existence of these  otherwise elusive Higgs particles.   
As  a    prototype example, we    calculate
diffractive production  cross-sections  of  the lightest   Higgs boson
within the framework of the Minimal Supersymmetric Standard Model with
explicit CP  violation.   Our   analysis shows that   the  challenging
regions of parameter space corresponding to a light CP-violating Higgs
boson might be accessible at the LHC provided suitable proton
tagging detectors are installed.

\newpage

Ultra-violet  completion  of  the  Standard Model  (SM)  suggests  the
existence of a new fundamental  scalar, known as the Higgs boson.  The
SM  Higgs  boson has  not  yet been  discovered  and  this provides  a
motivation  for  studies of  non-standard  Higgs  sectors.  A  natural
extension of  the SM  is to add  an additional Higgs  doublet, thereby
permitting  CP  violation in  the  Higgs  sector  of the  theory.   In
particular,  theories  based   on  supersymmetry  (SUSY)  require  the
existence  of  additional   Higgs  doublets  on  pure  field-theoretic
grounds~\cite{HPN}.   A~minimal realization  of  a softly-broken  SUSY
theory  is   the  so-called  Minimal   Supersymmetric  Standard  Model
(MSSM)~\cite{Haber}.   It is  known~\cite{APLB} that  third generation
squark  loops  can introduce  sizeable  CP  violation  into the  Higgs
potential of the MSSM.  Within this predictive CP-violating framework,
the  neutral Higgs  bosons will  mix  to produce  three physical  mass
eigenstates of indefinite CP  parity, labelled as $H_{1}$, $H_{2}$ and
$H_{3}$  in order  of increasing  mass~\cite{Pilaftsis:1999qt}.  After
the  inclusion of CP-violating  quantum effects,  each of  these three
mass eigenstates may have appreciable couplings to $W$ and $Z$ bosons.
A~salient feature of  this model is that the $H_1  ZZ$ coupling can be
significantly  suppressed,  reducing  the  LEP~II mass  limit  on  the
$H_{1}$ boson to as  low as 60 GeV~\cite{Carena:2001fw}.  In addition,
if  the two heavier  Higgs bosons  $H_{2,3}$ decay  predominantly into
pairs  of  the  lightest  Higgs boson  $H_1$~\cite{Choi:1999uk},  then
almost no lower  bound on the $H_1$ mass can be  derived at LEP, while
all three neutral Higgs particles may still remain hidden in the large
$b\bar{b}$ background at the Tevatron and LHC~\cite{Carena:2002bb}.

In this Letter we show that diffractive collisions at the
LHC may offer a unique  probe for   the   possible  existence of   light
CP-violating MSSM Higgs bosons, which could otherwise escape
detection in conventional  search channels.   To this
end, we calculate   production  cross-sections for  the   process $p+p
\rightarrow p+H_1+p$ in which  the final state consists  \textit{only}
of two intact protons and the decay products of the Higgs boson.

The  reaction $p+p \rightarrow p+H_1+p$  is often termed the exclusive
process~\cite{Schaefer,  Bialas,  Milana, Cudell:1995ki, Levin:1999qu,
Khoze:2000cy} and  has  both  experimental and  theoretical advantages
over the conventional non-diffractive  case.  If both outgoing protons
are tagged in   detectors a long   way downstream of the   interaction
point, a Higgs   mass resolution of  the  order of 1  GeV is  possible
\cite{DeRoeck:2002hk}   using   the  so-called  missing   mass method.
Furthermore,  because of the   requirement  that the outgoing  protons
remain intact and scatter  through small angles, only $0^{++}$ systems
can  be  produced.   This  has the   effect  of  suppressing  the  QCD
background     to     the     dominant     $b\bar{b}$      decay  mode
\cite{Khoze:2000mw,Khoze:2000jm}.   It has been  shown  that signal to
background ratios of order 3 could be achieved for a SM Higgs boson of
mass 120 GeV in the $b\bar{b}$ channel at the LHC, using a combination
of proton and $b$ jet tagging \cite{DeRoeck:2002hk}, for an integrated
luminosity of 30  fb$^{-1}$.  At the  Tevatron, with an estimated rate
of 0.2 fb \cite{Khoze:2001xm}, the situation is not so promising.  The
main reason for a such low cross-section is the size of the Higgs mass
relative  to  the total  center-of-mass    energy available for  Higgs
production.  In an   exclusive diffractive collision,  the  fractional
longitudinal momentum loss  of the  protons  $\xi $ is typically  less
than 10\%, and   therefore  the maximum  available  energy  for  Higgs
production  at  the  Tevatron   is  $\sqrt{\hat{s}}=\sqrt{\xi  _{1}\xi
_{2}s}$     $\sim$~200~GeV.  

In  stark contrast  to the  SM case,  the kinematic  situation changes
drastically for the case of  light CP-violating MSSM Higgs bosons with
masses smaller than 60 GeV.   As we will explicitly demonstrate below,
it may be possible to detect their production using exclusive 
diffractive collisions at the LHC.

Our  calculation of the  diffractive production  cross-section closely 
follows that  of \cite{Khoze:2001xm}.  
In particular,  we  consider the  Higgs  boson to  be
produced by gluon fusion, just as in the inclusive process, except that
an additional gluon is exchanged between the incoming protons in order
to  neutralize the  colour and  allow  the protons  to continue  their
journey  unscathed.  Appropriate suppression  factors are  included to
account for the probability not  to radiate any further particles into
the final state.

We write the cross-section for the exclusive diffractive production of
a colour singlet state, such as a CP-violating Higgs boson $H_i$ (with
$i=1,2,3$), as
\begin{equation}
\frac{\partial  \sigma}{\partial  y}\   =\ \int  
\frac{\partial^2\mathcal{L}(M^2,y)}{\partial M^{2} \partial y} 
\:\hat{\sigma }(M^{2})\, dM^{2},
\end{equation}
where $\hat{\sigma  }$  is the  cross-section for the  hard subprocess
producing  the        Higgs   (i.e.   $gg         \to  H_i$)       and
$\partial^2\mathcal{L}/\partial M^{2}  \partial y$ is the differential
luminosity  associated with Higgs  production at rapidity $y$ and mass
$M$.    For further  information    on  the  luminosity  we  refer  to
\cite{Khoze:2001xm} where all of the formulae we use are presented and
discussed   in  detail.  Here    we   focus on  the    hard scattering
cross-section.

The  amplitude for   resonant    Higgs production via gluon    fusion,
$gg\rightarrow H_{i}$, is~\cite{DM,Choi:1999aj,Choi:2001iu}
\begin{eqnarray}
  \label{higg}               \mathcal{M}_{ggH_{i}}\:               =\:
-\frac{M^2_{H_{i}}\alpha_s\,        \delta       ^{ab}}{4\pi\,       v
}\bigg[\,S^{g}_i(M_{H_{i}}^2)\bigg(  \epsilon_1\!\cdot\!  \epsilon_2\: -\:
\frac{2\,                                      k_1\!\cdot\!\epsilon_2\,
k_2\!\cdot\!\epsilon_1}{M_{H_{i}}^{2}}\,          \bigg)\:         -\:
P^{g}_i(M_{H_{i}}^2)\frac{2}{M_{H_{i}}^{2}}\langle  \epsilon  _{1}\epsilon
_{2}k_{1}k_{2}\rangle\, \bigg] ,
\end{eqnarray}
where  $v\simeq  246~{\textrm{GeV}}$, the   indices $a,b$ count  gluon
colours, $k_{1,2}$ and $\epsilon _{1,2}$  denote the four--momenta and
polarization  vectors of the  incoming gluons, and $\langle \epsilon_1
\epsilon _{2}k_{1}k_{2}\rangle  = \varepsilon_{\mu \nu \rho \sigma }\,
\epsilon _{1}^{\mu }\epsilon _{2}^{\nu }k_{1}^{\rho }k_{2}^{\sigma }$.
In addition, the  scalar  and pseudo-scalar form factors  $S^g_i$  and
$P^g_i$ are
\begin{eqnarray}
S_i^{g}(M_{H_{i}}^2)   &  =   &   \sum  _{f=b,t}   g_{H_{i}\bar{f}f}^{S}\,
F_{sf}\Big(M_{H_{i}}^{2}/4m_{f}^{2}\Big)\            -\           \sum
_{f=b,t}\,\sum_{j=1,2}
g_{H_{i}\tilde{f}_{j}^{*}\tilde{f}_{j}}\frac{v^2}{4m_{\tilde{f}_{j}}^{2}}
F_{0}\Big(M_{H_{i}}^{2}/4m_{\tilde{f}_j}^{2}\Big)\,,\nonumber\\
P^{g}_i(M_{H_{i}}^2)   &   =   &  \sum_{f=b,t}\,   g_{H_{i}\bar{f}f}^{P}\,
F_{pf}\Big( M_{H_{i}}^{2}/4m_f^{2}\Big)\, .
\label{eq:SP}
\end{eqnarray}
The loop   functions $F_{sf}(\tau)$, $F_{0}(\tau)$  and $F_{pf}(\tau)$
may be found in \cite{Choi:1999at}.  In the limit $\tau\rightarrow 0$,
$F_{sf}(0)=2/3$, $F_{0}(0)=1/3$  and  $F_{pf}(0)=1$.    The quantities
$g_{H_{i}\bar{f}f}^{S}$ and   $g_{H_{i}\bar{f}f}^{P}$  are the reduced
scalar and  pseudo-scalar couplings of  the Higgs  bosons $H_{i}$ to a
fermion $f$ in units of $g\, m_{f}/(2M_{W})$~\cite{Carena:2001fw}.  If
only   CP-violating    Higgs-mixing    effects    are      considered,
$g_{H_{i}\bar{f}f}^{S}$  and $g_{H_{i}\bar{f}f}^{P}$ assume the simple
forms
\begin{eqnarray} g_{H_{i}\bar{b}b}^{S} & = & O_{1i}/\cos \beta \, ,\qquad
g_{H_{i}\bar{b}b}^{P}\  =\   -\,O_{3i}\tan  \beta  \,   ,\nonumber  \\
g_{H_{i}\bar{t}t}^{S}   &   =   &   O_{2i}/\sin   \beta   \,   ,\qquad
g_{H_{i}\bar{t}t}^{P}\ =\ -\,O_{3i}\cot \beta \, ,
\end{eqnarray}
where $O_{\alpha i}$  is a 3-by-3  orthogonal matrix which relates the
weak eigenstates $\alpha =(\phi _{1},\phi _{2},a)=(1,2,3)$ to the mass
eigenstates  $i=(H_{1},H_{2},H_{3})=(1,2,3)$, with  $(\phi   _{1},\phi
_{2},a)^{T}=O_{\alpha \, i}(H_{1},H_{2},H_{3})^{T}$.  In our numerical
analysis, we also   take  into account  CP-violating  finite threshold
effects on the  effective Higgs--fermion--fermion couplings  generated
by    the   exchange     of     gluinos    and     charged   Higgsinos
\cite{Pilaftsis:2002fe,Carena:2002bb,key-4}.     Finally, the diagonal
couplings  of the    neutral    Higgs bosons   $H_{i}$  to   sfermions
$\tilde{f}_j$ are real and given by
\begin{equation}
v\,  g_{H_{i}\tilde{f}_{j}^{*}\tilde{f}_{j}}\ =\ \sum _{\alpha = \phi
 _{1},\phi _{2},a}\ \sum _{\beta ,\gamma = L,R}\left(\Gamma ^{\alpha
 \tilde{f}^{*}\tilde{f}}\right)_{\beta \gamma  }O_{\alpha   i}U_{\beta
 j}^{\tilde{f}*}U_{\gamma j}^{\tilde{f}}\, .
\end{equation} 
In the  above, $U^{\tilde{f}}$ are  the sfermion mixing  matrices that
relate     the      weak     to     mass      eigenstates:
$(\tilde{f}_{L},\tilde{f}_{R})^{T}          =          U^{\tilde{f}}\,
(\tilde{f}_{1},\tilde{f}_{2})^{T}$,   where   the  coupling   matrices
$\Gamma^{\alpha       \tilde{f}^{*}\tilde{f}}$      are      presented
in~\cite{Choi:1999aj,Carena:2001fw}.
\noindent
The amplitude (\ref{higg}) should be averaged over all gluon colours and
polarizations, leading to
\begin{equation}
  \label{eq:2} \overline{\mathcal{M}}_{ggH_i}  \ =\ \frac{M_{H_i}^2
\alpha _{s}}{4\pi v}\, S^g_i (M_{H_i}^2)~.
\end{equation}
\noindent
The corresponding cross-section for  $gg\to H_i$  is
\begin{equation}
\widehat{\sigma}^{\rm excl}(gg \rightarrow H_i\, ;M^{2})\ =\
K_i\,
\frac{\alpha_{s}^{2}}{16\pi v^2}\,
\Big|S^{g}_i(M_{H_i}^{2})\Big|^{2}\, \delta
\bigg(1-\frac{M_{H_i}^2}{M^2}\bigg)\,,
\label{eq:3}
\end{equation}
where    $K_i=1+\frac{\alpha_{s}(M_{H_i}^2)}{\pi}(\pi^2+\frac{11}{2})$
accounts for  virtual QCD corrections  \cite{Khoze:2001xm}. Note  that
this cross-section  is not  the same  as the  subprocess cross-section
which would be used in  inclusive production, the latter would include
an  additional factor of   $1/[2(N_c^2-1)]$, where $N_c   = 3$ is  the
number of colours.

\begin{figure}[ht]
\vspace{3.cm}
  \begin{picture}(140,350)(0,0)
    \put(0,0){\epsfig{file=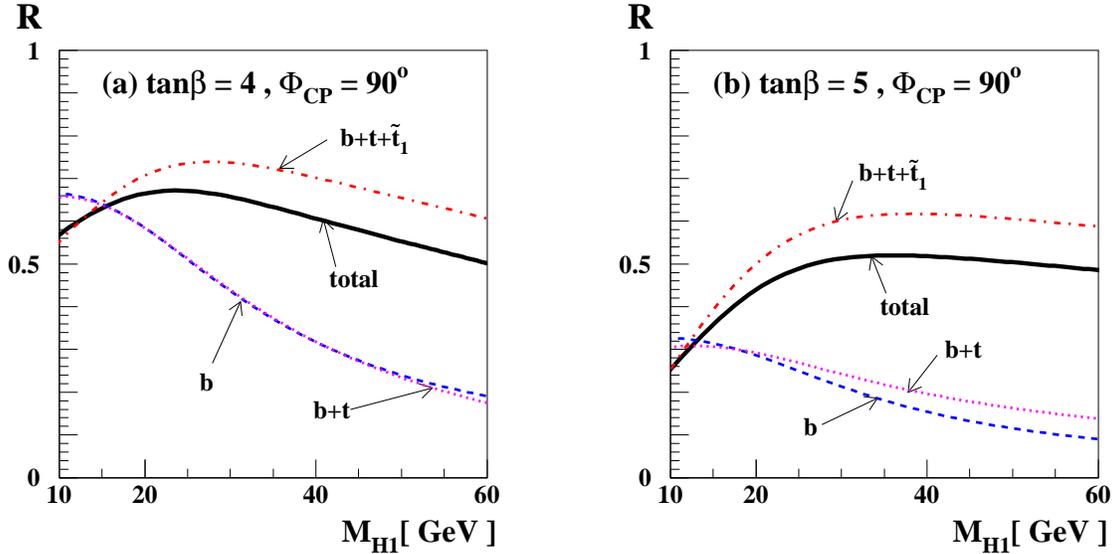,width=1.\textwidth}}
  \end{picture}
\vspace{-8cm}
\caption{\it The ratio $R\equiv |S_1^g(M_{H_1}^2)/S^g_{\rm
SM}(M_{H_{\rm SM}}^2)|$ assuming $M_{H_{\rm SM}}=M_{H_1}$ in the CPX
scenario \cite{Carena:2001fw} with $M_{\rm SUSY}=0.5$~TeV, $\Phi_{\rm
CP}=90^\circ$, and $(a)\,\tan\beta=4$ and $(b)\,\tan\beta=5$.}
\label{fig:ratio2}
\end{figure}

\begin{figure}[ht]
\vspace{-4cm}
  \begin{picture}(140,350)(0,0)
    \put(0,0){\epsfig{file=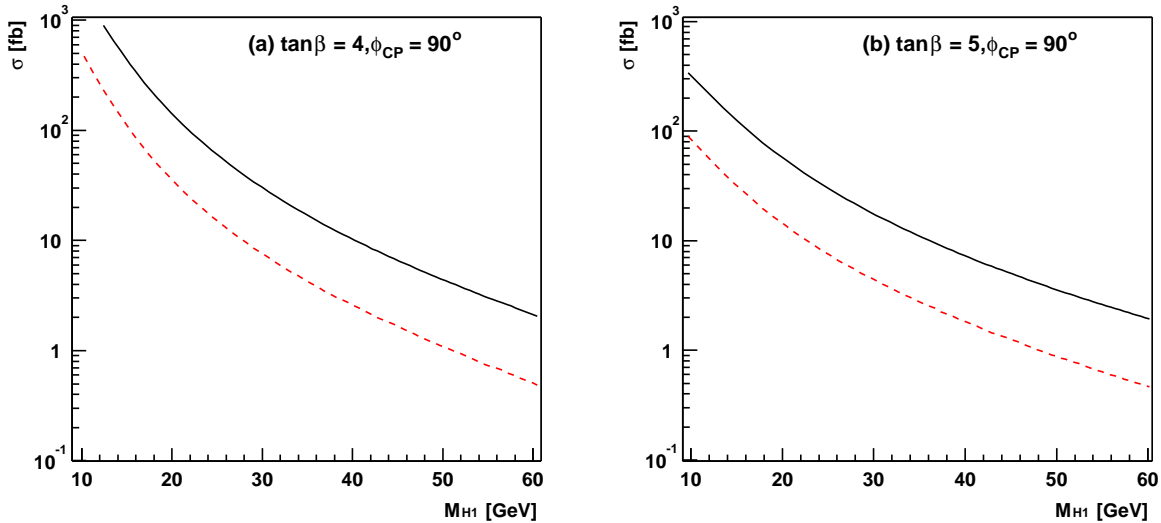,width=1.\textwidth}}
  \end{picture}
\caption{\it The cross-section for the process $p+p\rightarrow
p+H_1+p$ at the Tevatron (dashed line) and LHC (solid line) in the CPX
scenario \cite{Carena:2001fw} with $M_{\rm SUSY}=0.5$ TeV, $\Phi_{\rm
CP}=90^\circ$, and $(a)\,\tan\beta=4$ and $(b)\,\tan\beta=5$.  }
\label{xs}
\end{figure}
\noindent

For illustration, we consider the CPX scenario of \cite{Carena:2001fw}
with   $M_{\rm SUSY}   = 0.5$~TeV,  $\Phi_{\rm   CP}  = 90^\circ$  and
$\tan\beta = $ 4 and 5.  The phase $\Phi_{\rm CP}$ is defined as
\begin{equation}
\Phi_{\rm CP}\equiv
{\rm arg}(\mu A_t)=
{\rm arg}(\mu A_b)=
{\rm arg}(\mu\, {m_{\tilde{g}}} )\,.
\end{equation}
In  the    CPX     scenario,  moderate   values       of   $\tan\beta$
($3\stackrel{<}{{}_\sim} \tan\beta \stackrel{<}{{}_\sim} 6$) and large
CP     phases    $(90^{\circ}  \stackrel{<}{{}_\sim}   \Phi_{\rm   CP}
\stackrel{<}{{}_\sim} 120^{\circ})$  define a MSSM parameter space for
which   a    light    CP-violating   Higgs   boson,    with   $M_{H_1}
\stackrel{<}{{}_\sim}  50$~GeV,   cannot  be excluded  by  the  latest
analysis  of LEP2 data~\cite{Carena:2002bb}.  In  addition, it will be
difficult to observe such a light Higgs  boson in the channels: $(W/Z)
H_i  (\to bb)$ at the Tevatron,  and $gg\to H_i  (\to \gamma \gamma)$,
$t\bar{t} H_i  (\to b\bar{b})$ and  $WW \to H_i (\to \tau^+\tau^-)$ at
the LHC \cite{Carena:2002bb}.

Apart from the  aforementioned direct constraints  on the CPX scenario
based  on  standard  Higgs-search channels,  one   may worry about the
impact of indirect limits on  the large CP-violating phase  $\Phi_{\rm
CP}$ considered here that arise due to non-observation of electron and
neutron electric dipole  moments (EDMs).  However,  it has been  shown
recently~\cite{Pilaftsis:2002fe} that if the  first two  generation of
squarks    are heavier  than  about 3~TeV,    the  required degree  of
cancellation~\cite{EDM1}       between    the one-    and  higher-loop
contributions to EDMs due to large  gluino and third-generation squark
phases~\cite{EDM2}  is not   excessive.  In  particular,  for  low and
moderate values of $\tan\beta  \stackrel{<}{{}_\sim} 6$, the different
EDM contributions  may  add  destructively~\cite{Pilaftsis:2002fe} and
the   required degree   of   cancellation   is always smaller     than
60\%.\footnote{We   note   that   100\%    corresponds   to   complete
cancellation.}  Hence, we expect that   a full implementation of   EDM
constraints will not alter the  results of the  present analysis in  a
significant way.

In Fig.~\ref{fig:ratio2}, we show  the absolute value  of the ratio of
$S_1^g(M_{H_1}^2)$ for the lightest MSSM Higgs boson to that of the SM
Higgs boson with $M_{H_{\rm  SM}}=M_{H_1}$ as a function of $M_{H_1}$.
As is seen   in Eq.~(\ref{eq:SP}), the form factor  $S_1^g(M_{H_1}^2)$
can be decomposed  into contributions from the  bottom  quark, the top
quark  and the four  squarks.  The dashed  lines show the contribution
from the  bottom  quark,  the dotted  lines from   the  bottom and top
quarks,   and the  dash--dotted  lines from    the lighter top  squark
$\tilde{t}_1$ plus the   top and bottom  quarks.   Finally, the  solid
lines show   the  total contribution from   the  two quarks  and  four
squarks.  We note that $S^g_{\rm SM}(0)=4/3$ in the limit of vanishing
$M_{H_{\rm SM}}$.
For  small $\tan\beta$, the dominant contribution comes from  the
lighter  top squark $\tilde{t}_1$, the contributions  from
the other three heavier squarks being small and destructive.

In   Fig.~\ref{xs}, we  present  the exclusive
diffractive production   cross-section for the  lightest  neutral MSSM
Higgs boson. We use the same values of  
$\tan\beta$ and $\Phi_{\rm CP}$ as
in  Fig.~\ref{fig:ratio2}. Results are shown at both Tevatron 
(dashed line) and  LHC (solid line) energies. In both cases we
integrate over all $\xi  < 0.1$.  

Our results should be
compared    to   the  exclusive    diffractive   SM  Higgs  production
cross-section of $\sim$  3~fb at the LHC for  $M_H  =  120$ GeV 
\cite{DeRoeck:2002hk}.  The   conclusion    of
\cite{DeRoeck:2002hk} is  that a signal  to background ratio $\sim 3$
is possible for the 120 GeV Higgs, detected through its decay to
$b \bar{b}$.

For the lighter Higgs considered here, the signal to background ratio
will be reduced considerably since one anticipates a 
$S/B \propto \Gamma(H \to gg)/\Delta M \propto G_F M_H^3/\Delta M$ where
$\Delta M$ is the experimental resolution on the mass of the central
system (assumed to be 1 GeV in \cite{DeRoeck:2002hk}). 
There is a further suppresion since the $J_z=0$
selection rule becomes less effective at Higgs masses not much larger
than the $b \bar{b}$ threshold, i.e. the cross-section for exclusive
$b \bar{b}$ production is proportional to $m_b^2/M_H^2$.  
More detailed studies are clearly required before any definite conclusions 
can be drawn, however a rough 
estimate, based upon the calculation of the $b \bar{b}$ background
in \cite{DeRoeck:2002hk} in conjunction with the signal cross-sections
shown in Fig.\ref{xs}, suggests that a statistical significance 
($S/\sqrt{S+B}$) above $2 \sigma$ should
be obtainable at the LHC for Higgs masses above 20~GeV, provided  
suitable detectors are 
installed\footnote{The significance is aided by the fact
that the number of signal events rises rapidly with decreasing Higgs mass.}. 
Similar estimates indicate that the $b \bar{b}$ background will 
most likely be prohibitive at the Tevatron.


Although our focus  here has been  on light CP-violating Higgs bosons,
our study could be extended  to a number of  other scalar particles with
analogous phenomenological   features.   For  example,  light  radions
predicted    in  certain   higher-dimensional  scenarios  with  warped
geometry~\cite{radion,Carlos}  may couple significantly to gluons, but
feebly  to  $Z$ and   $W$  bosons,  thus  escaping  detection  in  the
conventional Higgs-search channels.

\section*{Acknowledgements}
We should like to thank Beate Heinemann, Valery Khoze and Misha Ryskin
for many helpful  discussions. BEC and JSL thank PPARC for
their support in funding part of this research.

\end{document}